\documentclass[10pt]{article}
\usepackage{subscript}
\usepackage{amsmath}
\usepackage{graphicx}
\usepackage{multicol}
\usepackage{float}
\usepackage{titlesec}
\usepackage{doi}
\usepackage{placeins}
\usepackage[numbers,sort&compress]{natbib}
\usepackage{booktabs}
\usepackage{makecell}
\usepackage{amssymb}

\usepackage[switch]{lineno}  
\usepackage{etoolbox} 

\titleformat{\section} 
  [hang]              
  {\centering\small\bfseries}   
  {}                  
  {0.5em}             
  {}                  

\usepackage[
  left=1in,
  right=0.8in,
  top=0.8in,
  bottom=0.8in
]{geometry}
\date{}

\usepackage[utf8]{inputenc}

\title{\textbf{Unraveling Spin Density Wave Order in Layered Nickelates $\mathrm{La_3Ni_2O_7}$ and $\mathrm{La_2PrNi_2O_7}$  via Neutron Diffraction}}

\author{
    Igor Plokhikh\textsuperscript{1,*}, \\
    Thomas J. Hicken\textsuperscript{1,\dag}, \\
    Lukas Keller\textsuperscript{1},  \\
    Vladimir Pomjakushin\textsuperscript{1}, \\
    Samuel H. Moody\textsuperscript{1}, \\
    Pascale Foury-Leylekian\textsuperscript{2},\\
    Jonas J. Krieger\textsuperscript{1} ,\\
    Hubertus Luetkens\textsuperscript{1},\\
    Zurab Guguchia\textsuperscript{1}, \\
    Rustem Khasanov\textsuperscript{1,\ddag}, \\
    and Dariusz Jakub Gawryluk\textsuperscript{1,\S}
}

\begin{document}
\maketitle

\noindent\textsuperscript{1} PSI Center for Neutron and Muon Sciences, Paul Scherrer Institute, Forschungsstrasse 111, 5232 Villigen, Switzerland  \\
\noindent\textsuperscript{2} Université Paris-Saclay, CNRS, Laboratoire de Physique des Solides, Orsay, 91405, France \\
\noindent\textsuperscript{*} Corresponding authors: \textsuperscript{*}\texttt{igor.plokhikh@psi.ch}, \textsuperscript{\dag}\texttt{thomas.hicken@psi.ch}, \textsuperscript{\ddag}\texttt{rustem.khasanov@psi.ch},\\ \textsuperscript{\S}\texttt{dariusz.gawryluk@psi.ch}

\begin{abstract}
The observation of pressure-induced superconductivity in two- and three-layer Ruddlesden-Popper nickelates has spurred intense interest in these materials as a platform for exploring unconventional superconductivity. While the ground state of these systems has been shown to exhibit magnetism, the direct determination of their magnetic structure remains elusive. This is a crucial aspect, as magnetism may play a role in the pairing mechanism of superconductivity in these materials. In this study, we resolve the magnetic structures in the bilayer (2222) polymorphs of {La$_3$Ni$_2$O$_7$} and {La$_2$PrNi$_2$O$_7$} compounds, using a combination of complementary techniques, namely neutron powder diffraction (NPD) and muon-spin rotation/relaxation ($\mu$SR). Magnetic neutron scattering in both samples emerges below $\sim$150~K and is observed at the ($q_x, \frac{1}{2}, 0$) position, with $q_x = 0$ and $\frac{1}{2}$ for La$_3$Ni$_2$O$_7$ and $q_x = 0$ for La$_2$PrNi$_2$O$_7$.  Alternating low- (0.05–0.075~$\mu_\text{B}$) and high- (0.66~$\mu_\text{B}$) magnetic moment stripes form a single layer; the bilayers are formed through antiferromagnetic stacking of single layers along the $c$-direction. The magnetic scattering with two propagation vectors $q_x = 0$ and $\frac{1}{2}$ in the undoped La$_3$Ni$_2$O$_7$ is attributed to two magnetic stacking polymorphs within a single crystallographic phase. The magnetic structures are substantiated by the  $\mu$SR spectra. These findings provide a detailed understanding of the magnetic ground state in bilayer nickelates, offering crucial insights into the possible precursor states that may influence the emergence of superconductivity in these materials.
\end{abstract}

\begin{multicols}{2}

\section*{I. INTRODUCTION}

\begin{figure}[H]
    \centering
    \includegraphics[width=0.85\columnwidth]{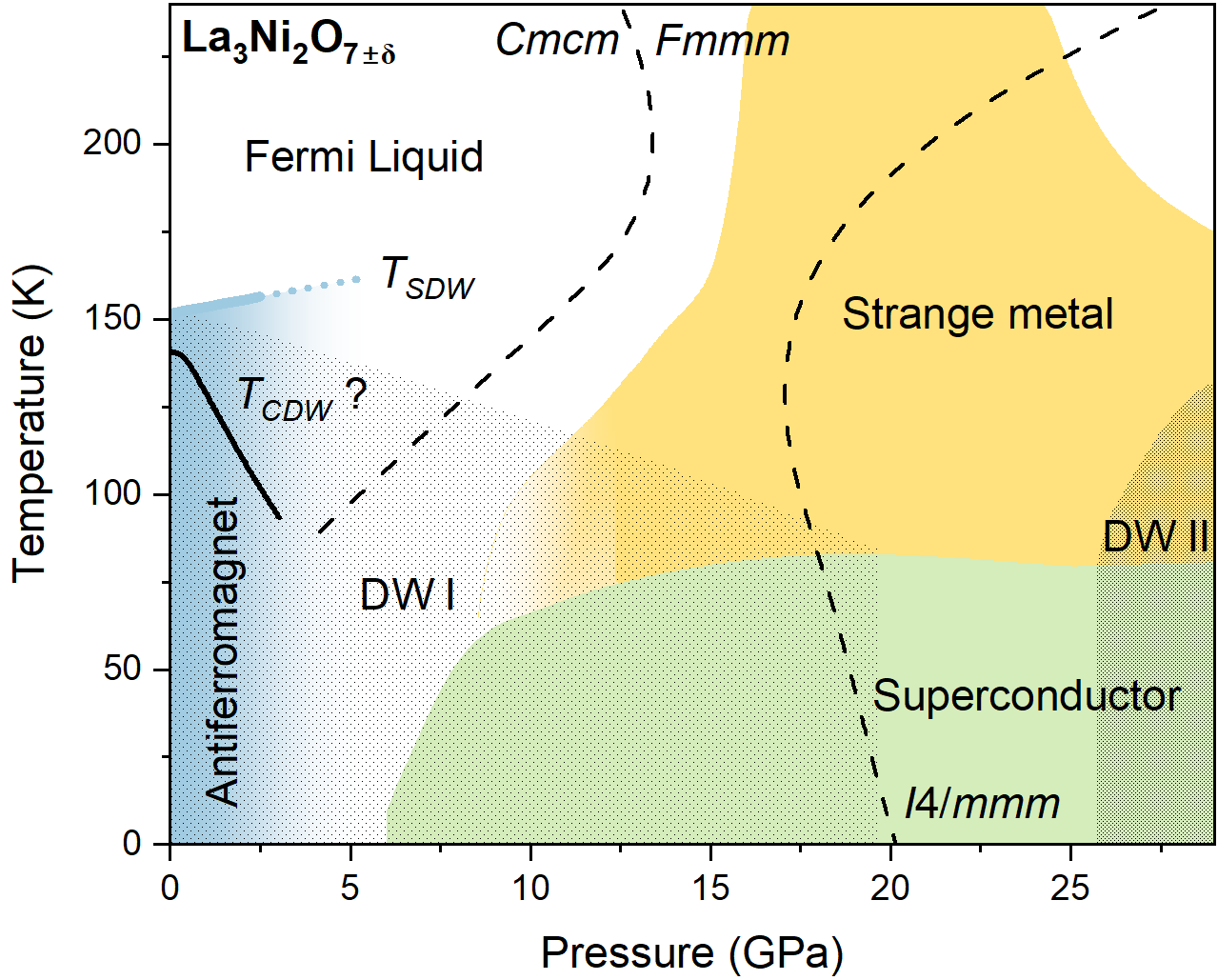}
    \caption{{\bf Temperature-pressure phase diagram of bulk La$_3$Ni$_2$O$_{7-\delta}$.} Solid and dashed lines represent the borders between different crystallographic phases; yellow, blues, green and shaded coloring represent the phases sorted by their physical properties. DW - density wave. Based on Refs. \cite{sun2023signatures, hou2023emergence, zhang2024high, meng2024density, wang2024pressure, zhang2024effects, khasanov2024pressure, wu2001magnetic, wang2024structure}. A more detailed version of this figure, including specific data points and references, is available in the Supplementary Information (SI). }
    \label{fig:PD}
\end{figure}

La$3$Ni$_2$O$_{7-\delta}$ and related layered nickelate systems have recently garnered significant attention due to their discovery as high-temperature superconductors under high pressure, marking this group one of the few transition-metal-oxide-based superconductors alongside the well-known copper- and iron-based families~\cite{wang2024normal, sun2023signatures, hou2023emergence, wang2024pressure, wang2024bulk, huang2024signature, sakakibara2024theoretical}.  This material's remarkable ability to exhibit superconductivity at temperatures exceeding 80~K under elevated pressure has opened a new frontier in the search for unconventional superconducting mechanisms. The complex phase diagram of {La$_3$Ni$_2$O$_{7-\delta}$}, analogous to that of cuprates and iron pnictides, which features an intricate interplay of electronic correlations, spin-(SDW) and multiple charge-(CDW) density waves, and pressure-induced structural transitions (Figs. \ref{fig:PD} and SI), renders it a fascinating field for further detailed explorations. This constitutes Ruddlesden–Popper-type (RP) nickelates a potentially significant subject in the broader context of understanding high-temperature superconductivity.

The RP layered structure, reminiscent of the structure found in cuprates, is a crucial component in establishing its emerging phenomena  \cite{anisimov1999electronic}. At ambient pressure in La$_3$Ni$_2$O$_{7-\delta}$ two orthorhombic crystallographic polymorphs - bilayer (2222) or alternating single-trilayer (1313) were reported \cite{wang2024long, chen2024polymorphism, PhysRevLett.133.146002}. Suppression of the orthorhombic distortion in the bilayer polymorph towards higher symmetry structures (\textit{Fmmm} or \textit{I}4/\textit{mmm}, SI) is a prerequisite for the onset of superconductivity~\cite{sun2023signatures, wang2024structure, PhysRevLett.133.146002}. The concurrent formation of the 1313 modification is proposed to be the reason for the filamentary and not bulk nature of superconductivity in La$_3$Ni$_2$O$_{7-\delta}$ \cite{wang2024bulk}. The formation of the 1313 phase can be suppressed by partial substitution of La with Pr, to form a La$_{3-x}$Pr$_x$Ni$_2$O$_{7-\delta}$ solid-solution series, with preserving the parent orthorhombic structure up to \textit{x} = 1. This substitution yields the samples exhibiting bulk superconductivity \cite{wang2024bulk}.

The complex nature of the strongly correlated La$_3$Ni$_2$O$_{7-\delta}$ has been investigated using a variety of experimental probes, including resonant inelastic X-ray scattering (RIXS)~\cite{chen2024electronic}, resonant soft x-ray scattering (RSXS) \cite{gupta2024anisotropic}, X-ray absorption spectroscopy \cite{Li2025}, ultrafast optical pump-probe spectroscopy \cite{meng2024density}, $\mu$SR~\cite{khasanov2024pressure, chen2024evidence}, and nuclear magnetic resonance (NMR)~\cite{kakoi2024multiband, dan2024spin}. The resonant X-ray studies have identified SDWs at low temperatures and ambient pressure, with an in-plane wave vector $(0.25, 0.25)$, hinting at complex magnetic interactions within the material \cite{chen2024electronic, gupta2024anisotropic}. However, the exact magnetic structure for La$_{3-x}$Pr$_x$Ni$_2$O$_{7-\delta}$  (\textit{x} = 0 and 1) has not been experimentally derived up to now. While high-pressure studies of La$_3$Ni$_2$O$_{7-\delta}$ have confirmed its superconducting phase, the connection between its magnetic and superconducting properties is still under investigation. So far, the actual ground state, the role of magnetic fluctuations, and their potential relationship with the superconducting pairing mechanism remain an open question \cite{tranquada1995evidence, guguchia2020using, Smidman2025}. Consequently, in the present study, the primary focus is directed towards the magnetic structure determination of the {La$_{3-x}$Pr$_x$Ni$_2$O$_7$} (\textit{x} = 0 and 1).

Although La$_3$Ni$_2$O$_{7-\delta}$ has been the subject of neutron scattering studies~\cite{xie2024strong, ling2000neutron, voronin2001neutron}, the precise nature of its magnetic order remained unclear up to now. In its sibling phase, trilayer homolog La$_4$Ni$_3$O$_{10}$, the magnetic order is detected through neutron scattering only in single crystals, as the magnetic scattering is substantially weaker than the nuclear~\cite{zhang2020intertwined}. Growing large high-quality single crystals of La$_3$Ni$_2$O$_{7-\delta}$ is challenging due to the coexistence of the 1313 and 2222 polymorphs~\cite{wang2024long, chen2024polymorphism, PhysRevLett.133.146002} complicated by concurrent epitaxial intergrowth with other homologs (La$_{n+1}$Ni$_n$O$_{3n+1}$, \textit{n} = 1, 3 and $\infty$).  As a result, determining the magnetic structure of the 2222 phase is both critical and difficult. In this study, using high-intensity neutron powder diffraction (NPD), we detect Bragg scattering linked to the long-range spin density wave in La$_{3-x}$Pr$_x$Ni$_2$O$_{7-\delta}$ (\textit{x} = 0 and 1) and propose spin-structure models consistent with both neutron and $\mu$SR data. The findings presented herein are, therefore, of significance for the elucidation of the electronic structure of {La$_3$Ni$_2$O$_{7-\delta}$} and its position within the broader family of transition-metal-oxide superconductors, thereby providing a foundation for further studies into its superconducting and magnetic behavior.

\section{II. RESULTS}
\subsection*{a. Basic characterization}

\begin{figure}[H]
    \centering
    \includegraphics[width=0.85\columnwidth]{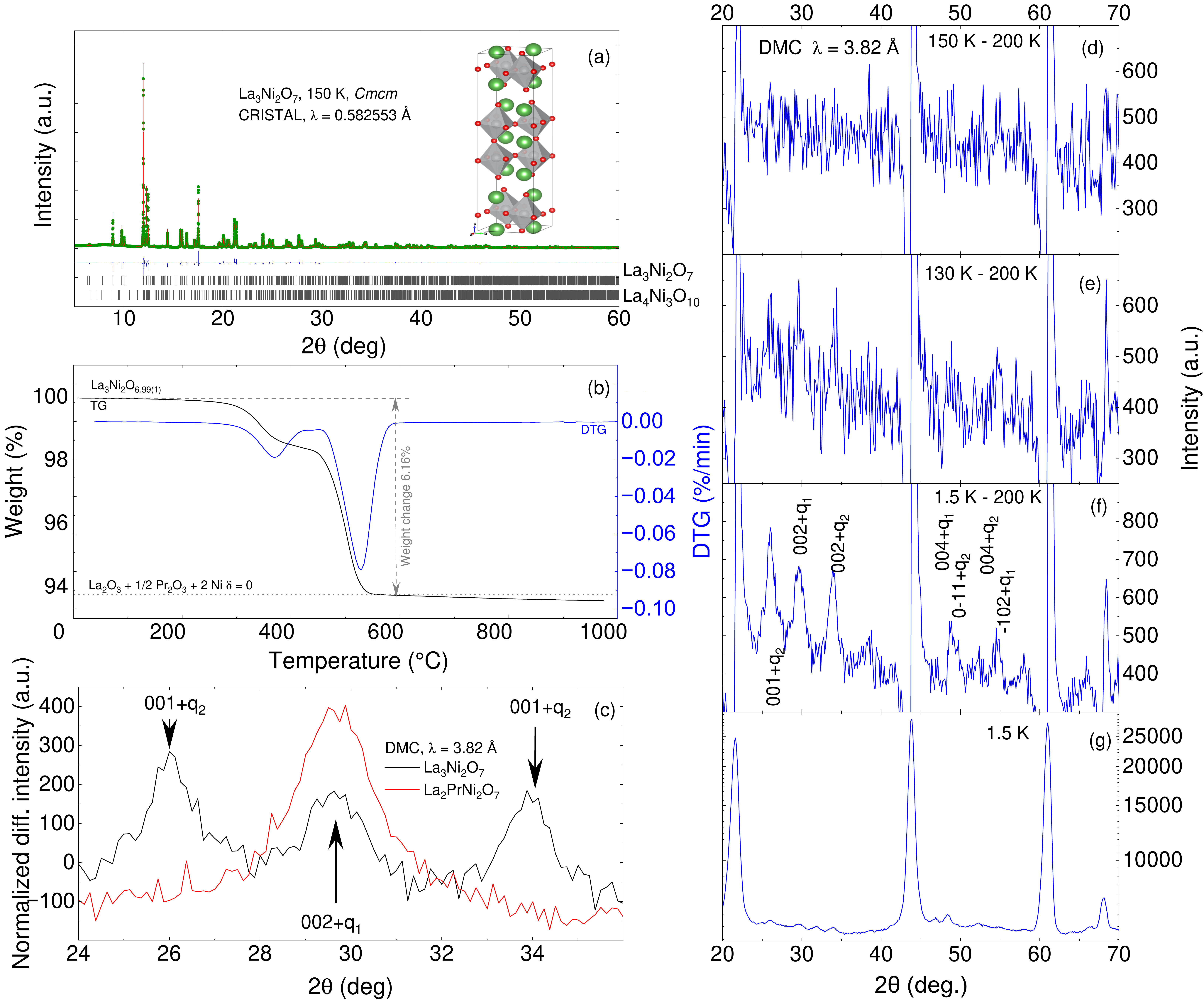}
    \caption{{\bf X-ray and neutron powder diffraction, and thermogravimetric analysis data of La$_3$Ni$_2$O$_7$ and La$_2$PrNi$_2$O$_7$.} (\textit{a}) Rietveld refinement of the x-ray powder diffraction patterns for La$_3$Ni$_2$O$_7$ measured at \textit{T} = 150 K. Green circles: observed data; red line: calculated pattern. The blue line represents the difference between measured and calculated data points. The vertical ticks indicate the positions of the Bragg reflections for the \textit{Cmcm}-{La$_3$Ni$_2$O$_7$} main phase and \textit{Cmce}-La$_4$Ni$_3$O$_{10}$ impurity. Insert shows projection of the refined crystal structure for La$_3$Ni$_2$O$_7$ as determined from the refinement. Grey, green, and red spheres stand for the Ni,  La, and O atoms, respectively. The structure is visualized using VESTA software \cite{momma2011vesta}  (\textit{b}) Hydrogen reduction thermogravimetric analysis for La$_3$Ni$_2$O$_7$: weight loss and derivative of the TG curve (DTG). (\textit{c}) Selected region of difference  neutron powder diffraction profiles for La$_3$Ni$_2$O$_7$ (1.5--200~K) and La$_2$PrNi$_2$O$_7$ (1.5--180~K). Neutron powder diffraction pattern for La$_3$Ni$_2$O$_7$ at (\textit{d}) \textit{T} = 150~K, (\textit{e}) \textit{T} = 130\,K, and (\textit{f}) \textit{T} = 1.5~K with \textit{T} = 200~K subtracted, and an as-measured pattern at (\textit{g}) \textit{T} = 1.5~K (\textit{y}-axis is in logarithmic scale). On the 1.5~K--200~K panel,  the indexes of magnetic reflections are indicated according to the indexing described in the text.}
    \label{fig:fig1}
\end{figure}

According to X-ray powder diffraction [Fig. \ref{fig:fig1}~(\textit{a}), SI], the polycrystalline samples of La$_3$Ni$_2$O$_{7-\delta}$ and La$_2$PrNi$_2$O$_{7-\delta}$  crystallize in the \textit{Cmcm} orthorhombic structure, \textit{i.e.}, they represent the 2222 polymorph. Synchrotron powder diffraction data for {La$_3$Ni$_2$O$_7$} reveal a small amount (\textit{ca.} 4\%) of La$_4$Ni$_3$O$_{10}$ impurity. Thermogravimetric analysis reveals $\delta$ = 0.01(1) and 0.04(1) for La$_3$Ni$_2$O$_{7-\delta}$ and La$_2$PrNi$_2$O$_{7-\delta}$ respectively, which indicates that the samples are nearly oxygen stoichiometric [Fig. \ref{fig:fig1} (\textit{b}), SI]; hereafter we will designate these samples as  {La$_3$Ni$_2$O$_7$} and {La$_2$PrNi$_2$O$_7$}.

\end{multicols}

\begin{multicols}{2}

Previously, in addition to the SDW, a second density wave (DW) transition was observed in the electrical transport data for undoped La$_3$Ni$_2$O$_7$ at \textit{T}  $\sim$130~K at ambient pressure \cite{khasanov2024pressure}. To check the possibility of long-range structural order, variable temperature synchrotron X-ray diffraction and high-resolution NPD experiments were performed. Patterns collected above and below the DW transition [SI] reveal no additional scattering in this region, opposing a long-range structural component associated with this transition. However, this does not entirely exclude the possibility of subtle structural distortion. Recent X-ray absorption spectroscopy studies have suggested the emergence of CDW below $\sim$150~K \cite{Li2025}. Nevertheless, variable temperature synchrotron X-ray diffraction measurements on high-quality single-crystals, similar to the study on {La$_4$Ni$_3$O$_{10}$} \cite{zhang2020intertwined} and total X-ray scattering with pair distribution analysis should be instrumental in addressing the structural component of the DW transition in {La$_3$Ni$_2$O$_7$}.

\subsection*{b. Neutron powder diffraction}

For La$_3$Ni$_2$O$_7$, the neutron powder diffraction pattern collected at DMC at
\textit{T} = 1.5~K (after subtracting the \textit{T} = 200~K data) reveals five additional reflections
at 2$\theta$ = 26.0, 29.7, 34.0, 48.9 and 54.8~deg. [Fig. \ref{fig:fig1} (\textit{f})]. Although weak,
these reflections can still be traced in the \textit{T }= 130~K  pattern
[Fig.~\ref{fig:fig1} (\textit{e})], whereas they are completely gone at \textit{T} = 150~K
[Fig.~\ref{fig:fig1} (\textit{d})], indicating a transition in this temperature range. These reflections are not broadened and their amplitude is on the order of 0.1 \% relative to the main nuclear reflections. 

\end{multicols}
\begin{figure}[H]
    \centering
    \includegraphics[width=0.9\columnwidth]{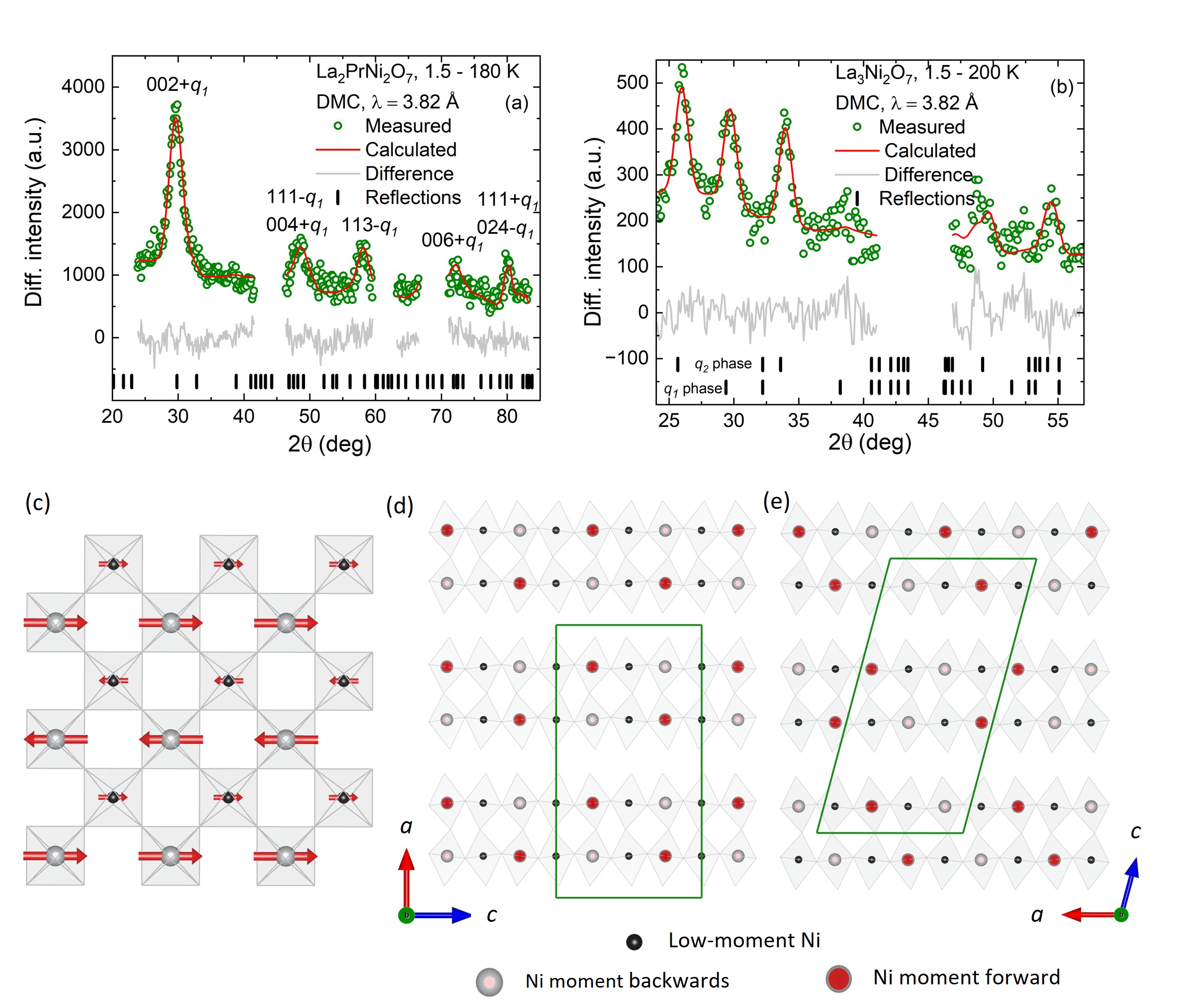}
    \caption{{\bf Neutron powder diffraction magnetic responses and magnetic models of La$_3$Ni$_2$O$_7$ and La$_2$PrNi$_2$O$_7$.}
    Rietveld refinement plots on difference neutron powder diffraction patterns for 1.5--180~K La$_2$PrNi$_2$O$_7$ (\textit{a}) and  1.5--200~K La$_3$Ni$_2$O$_7$ (\textit{b}). The gaps are at the positions of the principal nuclear reflections. (\textit{c}) Projections of arrangement of magnetic moments for a single Ni grid (view perpendicular to the single Ni-layer) for both \textit{q$_1$} and \textit{q$_2$} models. The magnetic moments on low- and high-moment sites are not to the scale. Stacking sequence of single Ni grids along the \textit{c}-axis for \textit{q$_1$} (\textit{d}) and \textit{q$_2$} (\textit{e}). The axes for (\textit{d}) and (\textit{e}) are given relative to the magnetic cells [Table \ref{tab:data}], which are outlined in green on the corresponding panels. 3D-views of magnetic unit cells are provided in SI. Panels (\textit{c}), (\textit{d}), and (\textit{e}) were visualized using VESTA software \cite{momma2011vesta}}
    \label{fig:fig2}
\end{figure}
\begin{multicols}{2}

As follows from a comparison of the difference neutron diffraction patterns for the pristine and Pr-substituted {La$_3$Ni$_2$O$_7$} [Fig. \ref{fig:fig1} (\textit{c})], the latter shows only one reflection at $\sim$ 29.7~deg. in the low-angle region. Following the findings of preceding $\mu$SR investigations ~\cite{khasanov2024pressure}, which suggested commensurate SDW order, all of the observed reflections in La$_2$PrNi$_2$O$_7$ can be indexed using a commensurate propagation vector \textit{q$_{1,t}$} = ($\tfrac{1}{4},\,\tfrac{1}{4},\,0)$ relative to the quasi-tetragonal (\textit{I}4\textit{/mmm}, SI) subcell, which transforms to \textit{q$_1$} = $(0,\tfrac{1}{2},0)$ in the orthorhombic \textit{Cmcm} basis (\textit{a} = 20.24~\AA, \textit{b} = 5.43~\AA, \textit{c} = 5.34~\AA). This is consistent with the wave vector deduced from previous RXS measurements \cite{chen2024electronic, gupta2024anisotropic}. Although the undoped La$_3$Ni$_2$O$_7$ exhibits notably more reflections emerging upon cooling, some of them (for instance, 2$\theta$ = 29.7~deg, Fig. \ref{fig:fig1} (\textit{c})) can still be indexed with \textit{q$_1$}, whereas others (for instance, 2$\theta$ = 26.0 and 34.0~deg.) can be indexed with \textit{q$_2$} = $\bigl(\tfrac{1}{2}, \tfrac{1}{2}, 0\bigr)$ (in \textit{I}4\textit{/mmm} setting -- \textit{q$_{2,t}$} = $\bigl(\tfrac{1}{4}, \tfrac{1}{4}, \tfrac{1}{2}\bigr)$), which is not symmetry equivalent to \textit{q$_1$}.

The neutron intensities of the reflections emerging at low temperatures decrease with increasing 2$\theta$ angle. This indicates that these reflections follow the magnetic form factor rather than the nuclear one \cite{furrer2009neutron}. Furthermore, as outlined in the previous section, the reflection corresponding to DW transition below 130 K are not detected by synchrotron diffraction data. Therefore, the reflections observed in NPD are judged to be due to magnetic order. Additionally, the two vectors (\textit{q\textbf{$_1$}} and \textit{q\textbf{$_2$}}) observed for the undoped sample are not related by the expected relation \textit{q$_{\mathrm{CDW}}$} = 2\textit{q$_{\mathrm{SDW}}$}, as is in the case of La$_4$Ni$_3$O$_{10}$~\cite{zhang2020intertwined}. Importantly, for La$_4$Ni$_3$O$_{10}$, the nuclear superstructure associated with the structural component of CDW is not visible in the neutron data either.

\subsection*{c. Magnetic models}
Symmetry-allowed models of spin structures were constructed and tested against the neutron data using the representation and magnetic symmetry routine implemented in \textsc{JANA2006}/\textsc{JANA2020} \cite{henriques2024analysis, petvrivcek2023jana2020} and ISODISTORT \cite{stokes_ISODISTORT, campbell2006isodisplace}. Group-subgroup trees for the maximal magnetic subgroups are provided in SI \cite{perez-mato2015symmetry,perez-mato2016symmetry}. A comparison of the most probable candidates for the \textit{q$_1$} (\textit{DT} point of Brillouin zone for the \textit{Cmcm} space group \cite{cracknell1979general, aroyo2014brillouin}) and
\textit{q$_2$} (\textit{S}-point of Brillouin zone) models, along with the corresponding Rietveld plots, is presented in Fig.~\ref{fig:fig2}. The model derived for the \textit{q$_1 $} in the case of La$_2$PrNi$_2$O$_7$ fits the \textit{q$_1$} intensities in La$_3$Ni$_2$O$_7$ equally well. The details of the magnetic structure for \textit{q$_1$} and \textit{q$_2$} are summarized in Table \ref{tab:data}.

\end{multicols}
\begin{table}[H]
    \centering
    \caption{Details of magnetic \textit{q$_1$} and \textit{q$_2$} structures in La$_3$Ni$_2$O$_7$: magnetic space groups \cite{perez2024guidelines, campbell2022introducing}, cell parameters, basis transformation from the parent nuclear \textit{Cmcm} cell to the magnetic cell, atomic positions, and corresponding magnetic moments. Only magnetic Ni atoms are provided.}
    
    \label{tab:data}
    \begin{tabular}{p{1.5cm} c c c c c c}
        \toprule
        \textbf{Atom} & \textit{x} & \textit{y} & \textit{z} & M$_x$ & M$_y$ & M$_z$ \\
        \midrule
        \multicolumn{7}{c}{\textbf{\textit{q$_1$} }model} \\
        \multicolumn{7}{c}{\textit{Pmmn.}1'$_\textit{c}$[\textit{Pmmn}] (UNI No. 59.413 \cite{campbell2022introducing})} \\ 
        \multicolumn{7}{c}{$a = 20.49~\text{\AA}, \quad b = 5.39~\text{\AA}, \quad c = 10.92~\text{\AA}$} \\
        \multicolumn{7}{c}{basis=$\{(1,0,0),(0,0,1),(0,-2,0)\}$} \\
        \midrule
        Ni1 & 0.847 & 0.25 & 0.749 & 0 & 0.87(3) & 0 \\
        Ni2 & 0.347 & 0.25 & 0.999 & 0 & 0.16(4) & 0 \\
        \midrule
        \multicolumn{7}{c}{\textbf{\textit{q$_2$}} model}\\ 
        \multicolumn{7}{c}{\textit{P}2$_1$/\textit{m}.1'\textit{$_a$}[\textit{P}2$_1$/\textit{m}] (UNI No. 11.55 \cite{campbell2022introducing})} \\ 
        \multicolumn{7}{c}{$a = 10.92~\text{\AA}, \quad b = 5.39~\text{\AA}, \quad c = 21.21~\text{\AA}, \quad \beta = 104.9~deg.$} \\ 
         \multicolumn{7}{c}{basis=$\{(0,2,0),(0,0,1),(1,-1,0)\}$} \\
        \midrule
        Ni1 & -0.075 & 0.25 & 0.597 & 0 & 0.87(3) & 0 \\
        Ni2 & -0.075 & 0.25 & 0.097 & 0 & 0.16(4) & 0 \\
        Ni3 & 0.327 & 0.25 & 0.403 & 0 & 0.16(4) & 0 \\
        Ni4 & 0.327 & 0.25 & -0.097 & 0 & 0.87(3) & 0 \\
        \bottomrule
    \end{tabular}
\end{table}

\begin{multicols}{2}

According to the derived models, for both \textit{q$_1$} and \textit{q$_2$} vectors,  a single magnetic Ni layer is formed by alternating high- and low-moment stripes. From the comparison of the stacking sequences along the \textit{c}-axis [Fig.~\ref{fig:fig2} (\textit{d}) and (\textit{e})], we can conclude that \textit{q$_1$} and \textit{q$_2$} correspond to two distinct stacking magnetic polymorphs. Assuming that the whole volume of La$_2$PrNi$_2$O$_7$ sample is occupied by the 2222 phase, and using thus derived scale factor, the values of the magnetic moment for the \textit{q$_1$} model in the La$_2$PrNi$_2$O$_7$ on high- and low-moment sites are $0.85(15)~\mu_{\mathrm{B}}$ and $0.15(5)~\mu_{\mathrm{B}}$. With this assumption applied to  La$_3$Ni$_2$O$_7$, considering that the sample contains two magnetic polymorphs (\textit{q$_1$} and \textit{q$_2$}) and constraining the values of low-and high-moments sites in \textit{q$_1$} and \textit{q$_2$} models, the derived magnetic moments are $0.87(3)\,\mu_{\mathrm{B}}$ and $0.16(4)\,\mu_{\mathrm{B}}$. Note, that the provided error bars are purely statistical from the least squares refinement; they do not account for errors introduced by the background correction. The latter is set up manually and not adjusted during the fitting. Hence, the physically realistic error bars should be larger. Later in this work, we test the neutron values of magnetic moments against the $\mu$SR data.

\end{multicols}

\begin{figure*}[htb]
    \centering
    \includegraphics[width=0.95\columnwidth]{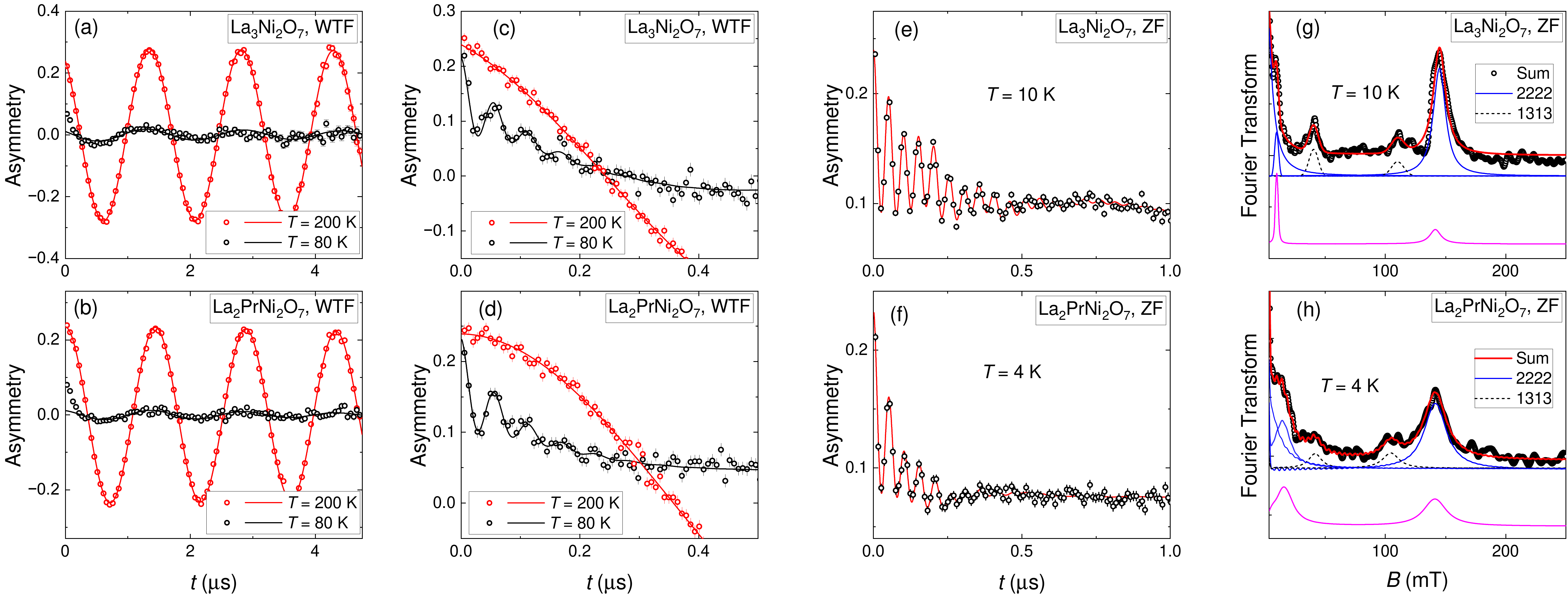}
    \caption{{\bf Magnetic response of La$_3$Ni$_2$O$_7$ and La$_2$PrNi$_2$O$_7$ probed by $\mu$SR.} (\textit{a})--(\textit{b}) $\mu$SR time-spectra collected in weak-transverse field experiments ($B_{\rm WTF}=5$~mT) above (\textit{T} = 200~K) and below (\textit{T} = 80~K) the SDW ordering temperature \textit{T$_N$}.
    (\textit{c})--(\textit{d})  The extended part of WTF-$\mu$SR data presented in panels ({\it a}) and ({\it b}).
    (\textit{e})--(\textit{f}) Zero-field $\mu$SR time spectra measured at low temperatures.
    (\textit{g})--(\textit{h}) Fourier transforms of the data presented in panels ({\it e}) and ({\it f}). The solid lines are fits by using Eqs. \eqref{eq:1} --  \eqref{eq:4} from the Methods section (see text for details). The pink lines are results of the dipolar-field calculations.}
    \label{fig:fig3}
\end{figure*}

\begin{multicols}{2}

\subsection*{d. $\mu$SR    Experiments}

The magnetic response of La$_3$Ni$_2$O$_7$ and La$_2$PrNi$_2$O$_7$ is further investigated by means of muon-spin rotation/relaxation ($\mu$SR). Two types of experiments were conducted: weak transverse field (WTF) $\mu$SR, with a weak magnetic field ($B_{\rm WTF}=5$~mT) applied perpendicular to the initial muon-spin polarization, and zero-field (ZF) $\mu$SR. The characteristic $\mu$SR spectra collected in ZF- and WTF-$\mu$SR experiments are presented in Fig. \ref{fig:fig3}.

In WTF experiments, the $\mu$SR time-spectra collected at \textit{T} = 200~K shows pronounced oscillations at $B_{\rm WTF}$ = 5~mT applied field with the initial asymmetry $A_0 \simeq 0.27$, in-line with characteristics of Flame and GPS $\mu$SR spectrometers \cite{FLAME_PSI, amato2017new}. At \textit{T} = 80~\textit{K} the initial asymmetry decreases down to $\sim 0.02$, suggesting the establishment of the magnetic state [Figs.~\ref{fig:fig3} ({\it a}) and ({\it b})]. This implies that magnetism in La$_3$Ni$_2$O$_7$ and La$_2$PrNi$_2$O$_7$ samples is representative of the bulk and reaches approximately 95\% already at
$T = 80$~K, suggesting that the systems are close to full order. A quick look at the extended part of the WTF data (up to $\simeq 0.5$~$\mu$s, Figs.~\ref{fig:fig3} ({\it c}) and ({\it d})) suggests that oscillations of the muon-spin polarization caused by the appearance of spin-density wave order are visible at 80~K and disappear at higher temperatures.

The high-statistic ZF-$\mu$SR responses are presented in  Figs. \ref{fig:fig3} ({\it f})--({\it g}). Panels ({\it f})/({\it e}) and ({\it h})/({\it g}) correspond to the ZF asymmetry spectra, representing the time evolution of the muon-spin polarization, and the Fourier transformation, showing the distribution of internal fields at $T=4$~K for La$_3$Ni$_2$O$_7$ and at $T=10$~K for La$_2$PrNi$_2$O$_7$, respectively.

\begin{figure}[H]
    \centering
    \includegraphics[width=0.95\columnwidth]{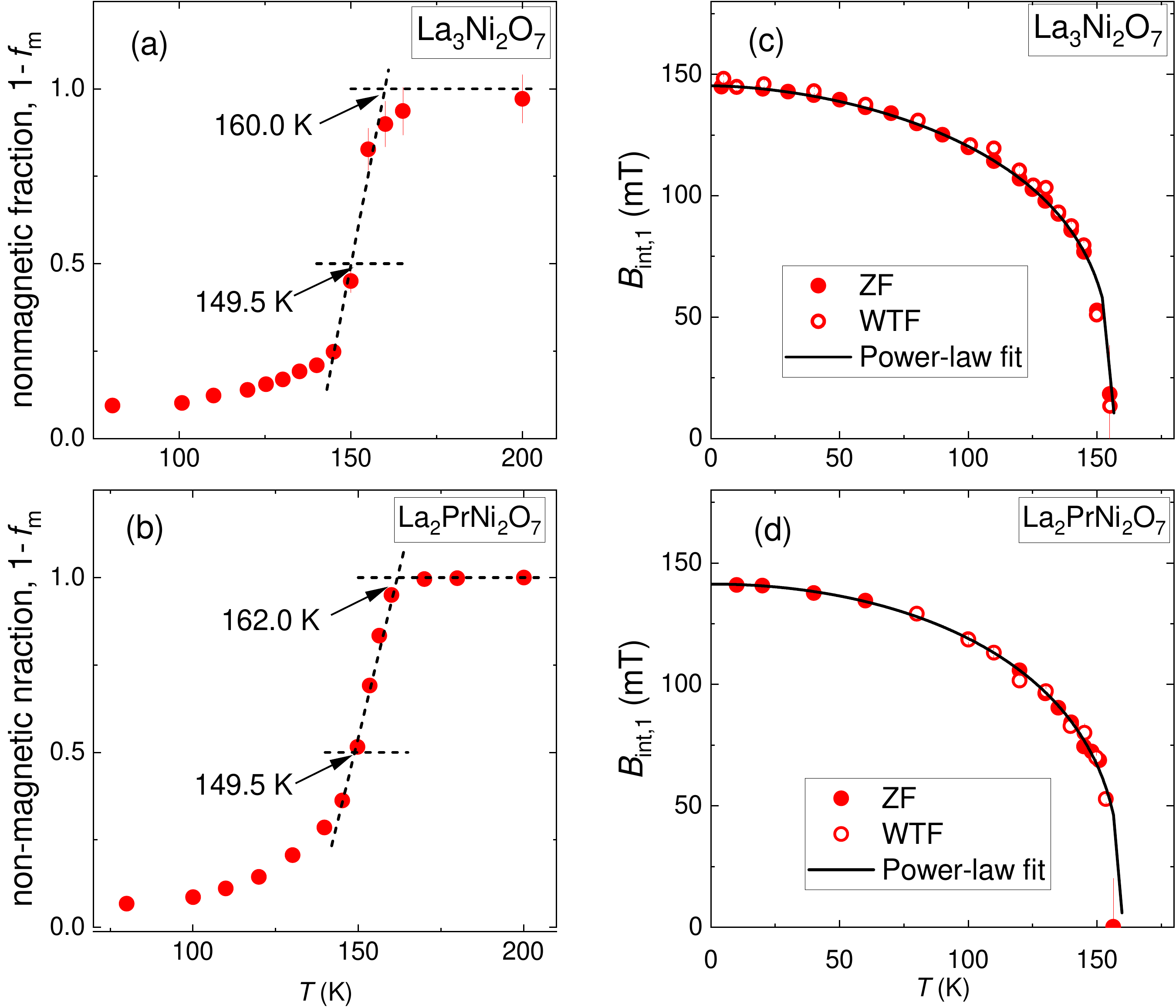}
    \caption{{\bf Magnetic fraction and internal field for La$_3$Ni$_2$O$_7$ and La$_2$PrNi$_2$O$_7$.}
    ({\it a})--({\it b}) Temperature dependence of the nonmagnetic volume fraction $1-f_m$ as obtained from fits of WTF $\mu$SR data. 
    ({\it c})--({\rm d}) Temperature dependence of fast precession component of the internal field $B_{int,1}$ as obtained from fits of WTF and ZF $\mu$SR data. The solid lines are power-law fits (Eq.~\ref{eq:power-law}) to the B$_{int,1}$(T) data. The displayed error bars for parameters are derived from fits of WTF and ZF-µSR data and correspond to one standard deviation from the $\chi^2$ 2 fits.}
    \label{fig:fig4}
\end{figure}

The WTF-$\mu$SR data were analyzed using Eqs. \eqref{eq:1}, \eqref{eq:2}, and~\eqref{eq:4} in the Methods section. The `long' and `short' time-window fits of the WTF data allow for the determination of the temperature evolutions of the nonmagnetic fraction ($1-f_{\rm m}$) and the strongest internal field component \(B_{\mathrm{int},1}\). The ZF~$\mu$SR data were analyzed using Eqs. \eqref{eq:1}, \eqref{eq:3}, and~\eqref{eq:4}.

Fits of high-statistics ZF-$\mu$SR data reveal that the ‘transversal’ part of Eq.\eqref{eq:3} consists of five terms [Figs.~\ref{fig:fig3} ({\it g}) and ({\it h})]. Three of these -- the fast-precessing, slow-precessing, and fast-relaxing components (denoted in blue) -- are similar to those reported in our previous study for the 2222 structural modification of La$_3$Ni$_2$O$_7$ \cite{khasanov2024pressure}. Two additional peaks at internal fields $B_{\rm int} \simeq 110$ and $40$~mT (denoted by dashed lines) may be associated with the 1313 polymorph \cite{khasanov2024lanio1313}.

A comparison of the corresponding line intensities indicates that in the La$_3$Ni$_2$O$_7$ sample, the 2222 phase occupies approximately 85–90\% of the total sample volume, while in La$_2$PrNi$_2$O$_7$, the 2222 phase accounts for more than 90\% of the volume. The intensity ratios and the values of the fast-precessing ($B_{\rm int,1}$) and slow-precessing ($B_{\rm int,2}$) internal field components associated with the 2222 structural phase were found to be 92\%/8\%, $B_{\rm int,1} \simeq 145$~mT and $B_{\rm int,2} \simeq 10$~mT for La$_3$Ni$_2$O$_7$, and 82\%/18\%, $B_{\rm int,1} \simeq 143$~mT and $B_{\rm int,2} \simeq 14$~mT for La$_2$PrNi$_2$O$_7$ respectively.

The results of the data analysis of WTF and ZF-$\mu$SR experiments are presented in Fig.~\ref{fig:fig4}. Panels ({\it a}) and ({\it b}) show the temperature evolution of the non-magnetic volume fraction $\bigl(1 - f_{\mathrm{m}}\bigr)$ as obtained in WTF-$\mu$SR measurements of La$_3$Ni$_2$O$_7$ and La$_2$PrNi$_2$O$_7$ samples, respectively. In both systems, the initial asymmetry recovers above $\sim$ 160 K, which allows determination of the mid ($T_{N}^{\mathrm{mid}}$) and the onset  ($T_{N}^{\mathrm{onset}}$)  temperatures of the magnetic transitions. The corresponding transition temperatures $\bigl(T_{N}^{\mathrm{mid}} / T_{N}^{\mathrm{onset}}\bigr)$ were found to be $149.5\,\mathrm{K} / 160.0\,\mathrm{K}$ and $149.5\,\mathrm{K} / 162.0\,\mathrm{K}$ for La$_3$Ni$_2$O$_7$ and La$_2$PrNi$_2$O$_7$.

The temperature evolution of the most pronounced fast-precessing internal field component $B_{\mathrm{int},1}$ is presented in Figs.~\ref{fig:fig4}~({\it c}) and ({\it d}). The solid lines are fits of a phenomenological power-law function
\begin{equation}
B_{\rm int}(T) = B_{\rm int}(0)\left[1-(T/T_{\rm N})^\alpha \right]^\beta
 \label{eq:power-law}
\end{equation}
to $B_{\mathrm{int},1}(T)$. Here $B_{\rm int}(0)$ is the zero-temperature value of the internal field, $T_{\rm N}$ is the N\'{e}el temperature and $\alpha$ and $\beta$ are the power exponents. The corresponding values of $T_{\rm N}$, $B_{\mathrm{int}}(0)$, $\alpha$, and $\beta$ are 156.5(1.1)~K, 145.2(3)~mT, 1.73(7), and 0.31(2) for La$_3$Ni$_2$O$_{7}$, and  159.8(1.2)~K, 141.3(5)~mT, 2.02(15), and 0.35(3) for La$_2$PrNi$_2$O$_7$, respectively.

The results of WTF- and ZF-$\mu$SR experiments reveal a nearly identical magnetic response in both La$_3$Ni$_2$O$_7$ and La$_2$PrNi$_2$O$_7$ samples. The distributions of internal magnetic fields, as well as their low-temperature values, remain closely similar. Most parameters, including magnetic transition temperatures $T_{\rm N}$, $T_{\rm m}^{\rm mid}$, and $T_{\rm m}^{\rm onset}$, as well as power-law exponents $\alpha$ and $\beta$, were found to be nearly identical in both compounds.

\subsection*{e. Comparison of NPD and $\mu$SR Magnetic Responses: Consistent Trends.}

\begin{figure}[H]
    \centering
    \includegraphics[width=0.95\columnwidth]{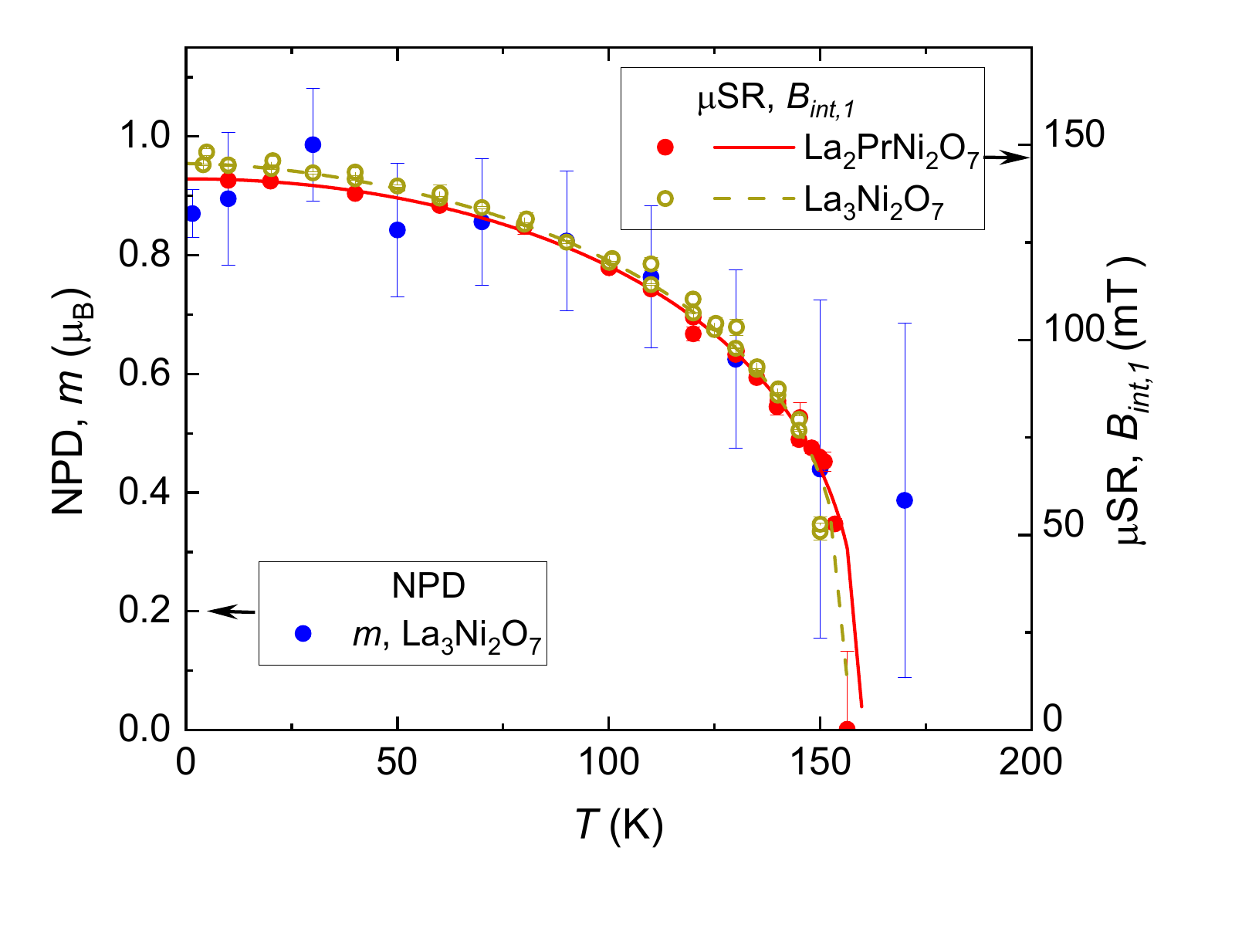}
    \caption{{\bf Comparison of NPD and $\mu$SR magnetic responses.}
    Comparison of the temperature dependence of magnetic moment (\textit{m}, $\mu_B$) for the high-moment site in {La$_3$Ni$_2$O$_7$} derived from NPD data according to the models for \textit{q$_1$} and \textit{q$_2$} with the temperature evolution of fast precession component of the internal field ($B_{\rm int,1}$, mT) derived from $\mu$SR measurements for La$_3$Ni$_2$O$_7$ (dashed line) and La$_2$PrNi$_2$O$_7$ (solid line). The error bars for the neutron and muon data are statistical.}
    \label{fig:fig5}
\end{figure}

In order to test that the observed magnetic scattering at low temperature (below \textit{T} $\approx$ 150~K) is common with SDW observed by $\mu$SR experiments, the order parameter evolution as the function of temperature is presented in Fig.~\ref{fig:fig5}, superimposing the results of both NPD (La$_3$Ni$_2$O$_7$) and $\mu$SR (La$_{3-x}$Pr$_x$Ni$_2$O$_7$, \textit{x} = 0 and 1). The neutron and muon data show a clear agreement, although the error bars for neutron-derived magnetic moments are relatively large compared to the magnetic order parameter as observed by $\mu$SR.

To test the feasibility of the neutron-derived magnetic structures to describe the $\mu$SR response, we have performed simulations of the dipole field at the DFT-calculated muon site, following the approach in Ref.~\cite{khasanov2024pressure}. Both La$_3$Ni$_2$O$_7$ and La$_2$PrNi$_2$O$_7$ exhibit qualitatively similar ZF-$\mu$SR responses [Figs.~\ref{eq:3}~({\it f})--({\it g})], suggesting the necessity for some sites to have zero magnetic moment to explain the low-field muon site. Our neutron results suggest, instead, that there are sites with a very low moment ($\lesssim 0.15$~$\mu_{\rm B}$), rather than zero.

Dipole field simulations for the \textbf{$q_1$} model [Figs.~\ref{fig:fig2} ({\it c}) and ({\it d})] suggest that this spin structure is compatible with the $\mu$SR data for La$_2$PrNi$_2$O$_7$ when the high and low moments are 0.655~$\mu_{\rm B}$ and 0.075~$\mu_{\rm B}$, respectively. With these magnetic moments, the simulations produce a large internal field of $\simeq$ 141~mT, and a lower field of $\simeq 14$~mT, consistent with the experimental data ($B_{\rm int,1} \simeq 143$~mT and $B_{\rm int,2} \simeq$ 14~mT). The pink line in Fig.~\ref{fig:fig3}({\it g}) represents the simulated field distribution. From the measurements, we expect 82\%/18\% of the muons to stop in the high/low field sites respectively, however the simulations suggest that there are an equal number of muons stopping in both fields. In the simulation, we assume a perfectly ordered infinite magnetic structure; the finite correlation length in the real system will change the fraction of muons stopping in each type of site. The significant difference between the simulation and experimental data may suggest that the correlation length is rather low at least in one dimension in La$_2$PrNi$_2$O$_7$.

We now turn to the undoped La$_3$Ni$_2$O$_7$, which, unlike the Pr substituted sample, shows magnetic neutron scattering at two vectors, $q_1$, and $q_2$, with \textit{q$_2$} being dominant.  The $q_2$ model provided in Figs.~\ref{fig:fig2}~({\it c}) and ({\it e}) is compatible with the muon data when the high and low moments are  0.66~$\mu_{\rm B}$  and 0.05~$\mu_{\rm B}$ respectively, with an equal fraction of the muons sensitive to an internal field of around 10~mT or 141~mT. The values of the internal field are in agreement with the results presented in Fig.~\ref{fig:fig3}~({\it h}) and in Ref.~\cite{khasanov2024pressure}. Once again, experimentally fewer muons stop in the low-field site compared to the simulation, which again likely reflects a relatively short correlation length.

The differences between the values of magnetic moments derived from NPD and $\mu$SR are $\sim$ 0.1$\,\mu_{\mathrm{B}}$ and $\sim$ 0.2$\,\mu_{\mathrm{B}}$ for low- and high-moment sites, \textit{e.g.} mostly fall close or slightly above the error bars calculated from the Rietveld fit of the neutron data. This further validates the models proposed from the representation and magnetic symmetry analysis. Still, the muon values are systematically lower than the neutron ones. We see several reasons for this discrepancy.  First, as mentioned above, the magnetic signal seen by neutrons is rather weak, which causes an arbitrary process of background correction to introduce sizable uncertainties. Another source of systematic error is the presence of 10 to 15~\% of 1313 polymorph, as evident from the $\mu$SR data but not from the neutron and X-ray diffraction. Overall, we judge the representation and magnetic symmetry analysis done on neutron data to directly yield the spin arrangement, whereas $\mu$SR is more faithful in determining the absolute values of magnetic moments.

\section{III. CONCLUSIONS AND OUTLOOK}
Previous neutron studies of La$_3$Ni$_2$O$_{7-\delta}$ revealed no elastic scattering (Bragg diffraction) associated with long-range magnetic order \cite{wang2024pressure, xie2024strong, ling2000neutron}, whereas resonant X-ray scattering \cite{chen2024electronic, gupta2024anisotropic} and $\mu$SR \cite{khasanov2024pressure, chen2024evidence} experiments, in line with macroscopic property measurements, show the presence of SDW. The lack of magnetic diffraction in neutron experiments is ascribed either to a small magnetic moment or to the short-range character of the order \cite{xie2024strong, khasanov2024pressure}. Contrary to that, the neutron data presented in our report unambiguously indicate the long-range magnetic order evident from the magnetic Bragg peaks [Figs. \ref{fig:fig1} and \ref{fig:fig2}], associated with a magnetic moment of $\sim$ 0.66~$\mu_{\mathrm{B}}$ on half of the Ni atoms.

This apparent contradiction between the previous reports and our current work can be ascribed to several factors related to the sample and the neutron experiment setup. Our samples are close to ideal stoichiometry. Theory \cite{ni2025spin} and $\mu$SR \cite{chen2024evolution} studies suggested oxygen stoichiometry as the crucial factor defining the ground state magnetic order in the title material.  Additionally, our measurements were performed on large samples ($\sim$ 10 g) using a recently upgraded high-flux neutron diffractometer dedicated to studying small magnetic moment systems \cite{PSI_DMC}.

One peculiar feature of magnetic order in La$_3$Ni$_2$O$_7$ is scattering at two magnetic propagation vectors, \textit{q$_1$} and \textit{q$_2$}. The absence of scattering at the \textit{q$_2$} vector in the Pr-substituted sample [Fig. \ref{fig:fig1} (\textit{c})], which contains comparable amounts of the 1313 admixture, further validates our conclusion about two-vector indexing in the pristine sample. For this reason, we oppose the possibility of double-propagation-vector-scattering being due to the presence of 1313 polymorph. Comparing the spin structures in both samples, we can conclude that the magnetic order pattern in single Ni-layers for both structures is identical. In contrast, the difference is in the stacking sequence along the \textit{c}-axis. This might indicate a quasi-2D character of magnetic order in {La$_3$Ni$_2$O$_7$}, with significant role of stacking faults and local inhomogeneities defining actual ground-state magnetic order \cite{zhou2024revealing}. This aligns with the fact that the Pr-doped sample features both improved microstructure \cite{wang2024bulk} and just magnetic scattering at a single propagation vector [Fig. \ref{fig:fig1} (\textit{c})]. A very similar phenomenon has been observed in other layered compounds  \cite{plokhikh2022competing, plokhikh2023magnetic, plokhikh2023magneticGD, sun2025resolving, akatsuka2024non}.   Theory studies should shed light on the origin of the observed magnetic polymorphism.

This study presents the first direct observation of a spin-density wave (SDW) state in bilayer nickelates using neutron diffraction, providing key insights into the magnetic ground state of these materials. Previous $\mu$SR studies suggested that the SDW order is suppressed under applied pressure \cite{khasanov2024pressure}, raising the question of its connection to the emergence of superconductivity. Future high-pressure neutron diffraction experiments will be essential to determine whether the suppression of this magnetic order is directly linked to the onset of superconductivity. If confirmed, this would establish the SDW as a competing instability and a possible precursor state for unconventional superconductivity in these systems.

\section*{IV. METHODS}

\subsection*{a. Sample preparation}

The samples studied in this work were prepared via a solid-state reaction
from NiO (4N8, Alfa Aesar), La$_2$O$_3$ (5N, Sigma-Aldrich), and Pr$_6$O$_{11}$ (5N, Sigma-Aldrich) through a mechanochemically-assisted
process similar to that described previously Ref.~\cite{khasanov2024pressure}.
Stoichiometric mixtures of the calcinated oxides were ball-milled (800~rpm, 5~min grinding, 5~min pause, 20~cycles), pressed
into pellets, and annealed in oxygen flow for 100~h at
1250~$^\circ\mathrm{C}$  (La$_3$Ni$_2$O$_7$) or for 70~h at 1100~$^\circ\mathrm{C}$ (La$_2$PrNi$_2$O$_7$) with intermediate grindings and a post-annealing stage (24~h/500~$^\circ\mathrm{C}$). For the La$_2$PrNi$_2$O$_7$, the annealing temperature has been chosen according to Ref. \cite{wang2024bulk}.

\subsection*{b. X-ray powder diffraction}
Samples' purity is checked, and data for structure refinement are collected at room temperature using a \textsc{Bruker} \textsc{AXS D8} \textsc{Advance} diffractometer (CuK$\alpha$ radiation). Variable temperature synchrotron powder diffraction datasets were collected at the CRISTAL beamline (Soleil Synchrotron) using $\lambda$ = 0.582553 \AA\ radiation. A Rietveld analysis of the obtained x-ray diffraction pattern at 150 K is performed utilizing the JANA software \cite{petvrivcek2023jana2020}. This analysis yielded the following unit cell parameters of the orthorhombic structure (\textit{Cmcm}, Space Group No.~63): \textit{a} = 20.4970(1); \textit{b} = 5.44781(3); \textit{c} = 5.37835(3)~Å. The reliability factors of the fit are as follows: \textit{R$_p$} = 5.7; \textit{R$_{WP}$} = 7.6; GOF = 9.7; \textit{R$_I$} = 4.1.

\subsection*{c. Oxygen content determination}
The oxygen content in the La$_{3-x}$Pr$_x$Ni$_2$O$_{7-\delta}$ (\textit{x} = 0 and 1) samples was determined by hydrogen-reduction thermogravimetric analysis (TGA). A \textsc{Netzsch~STA~449F1} simultaneous thermal analyzer (STA) was used for the TGA experiments. The initial sample masses were 144.04~mg and 83.29~mg for \textit{x} = 0 and 1, respectively. The gas utilized was a mixture of 5~vol\% hydrogen (\textsc{Messer Schweiz AG}, 5N) in helium (\textsc{PanGas}, 6N), with a flow rate of 70~ml/min. The sample was heated from room temperature to 1000$^\circ\mathrm{C}$ at a rate of  1$^\circ\mathrm{C/min}$ and subsequently cooled to room temperature.  In order to minimize buoyancy effects and friction forces resulting from the vertical gas flow, and thereby ensure an accurate estimation of weight changes, a baseline measurement was performed under identical conditions before measuring the actual samples. The total weight loss can be attributed to the reduction of La$_{3-x}$Pr$_x$Ni$_2$O$_{7 -\delta}$ to metallic nickel (Ni) and \textit{R}$_2$O$_3$ (\textit{R} = La or Pr). The residual weight was estimated at a temperature corresponding to the reaction termination ($T_{\mathrm{f}}$), determined from the final plateau in the first derivative of the thermogravimetric (TG) weight-loss curve.

\subsection*{d. Neutron powder diffraction}
High-intensity NPD datasets were collected on the cold-neutron powder diffractometer \textsc{DMC} \cite{PSI_DMC, schefer1990versatile} with 3.82~\AA\ neutrons and high-resolution NPD dataset at the thermal neutron HRPT diffractometer \cite{fischer2000high} with 1.89~\AA\ neutrons, both located at Paul Scherrer Institute. $\sim$10~g of La$_3$Ni$_2$O$_7$ and $\sim$11~g of La$_2$PrNi$_2$O$_7$ samples were loaded into V-cans, placed in an Orange He-cryostat, and measured down to the base temperature of \textit{ca.} 1.5 - 1.8~K. Analysis of the diffraction datasets,  including representation and magnetic symmetry, was performed using \textsc{Jana2006}/\textsc{Jana2020} software according to the standard mathematical approach for powder diffraction data (pseudo-Voigt profile function and manual background) as described in Refs. \cite{henriques2024analysis, petvrivcek2023jana2020}. The group-subgroup relations were analyzed using ISORDISTORT \cite{campbell2006isodisplace, stokes_ISODISTORT} and k-SUBGROUPSMAG software of the Bilbao Crystallographic Server \cite{perez-mato2015symmetry, perez-mato2016symmetry}.

\subsection*{e. Muon-spin rotation/relaxation experiments}
Zero-field (ZF) and weak transverse-field (WTF) $\mu$SR experiments were carried out at the Paul Scherrer Institute (\textsc{PSI}), Switzerland. Experiments were performed at the $\pi E3$ beamline using \textsc{GPS}~\cite{amato2017new} and \textsc{Flame} \cite{FLAME_PSI} muon spectrometers. The $\mu$SR data were analyzed using the \textsc{Musrfit}  software package \cite{suter2012musrfit}.  

\subsection*{f. Fits of WTF- and ZF-$\mu$SR data}
The fit to the experimental WTF- and ZF-$\mu$SR data was performed using the following functional form:
\begin{equation}
A(t) = A_0 \, P(t)
     = A_0 \bigl[f_m \, P_m(t) + \bigl(1 - f_m\bigr)\,P_{\mathrm{nm}}(t)\bigr].
     \label{eq:1}
\end{equation}
Here, $A_0$ is the initial asymmetry and $P(t)$ represents the time evolution of the muon-spin polarization. The polarization function $P(t)$ is further divided into the magnetic (m) and nonmagnetic (nm) contributions with weights $f_m$ and $1 - f_m$, respectively. The magnetic part is caused by the appearance of spin-density wave (SDW) order and is described as:

\begin{equation}
P_{m}(t) = \frac{2}{3}\sum_i f_i  e^{-\lambda_{{\rm T},i}t} \cos \left(\gamma_{\mu} B_{{\rm int,}i}t \right)
              + \frac{1}{3} e^{-\lambda_{{\rm L}} t}.
              \label{eq:2}
\end{equation}
Here, $\gamma_{\mu} = 851.616\,\mathrm{MHz/T}$ is the muon gyromagnetic ratio, $\lambda$'s denote the exponential relaxation rates, and $f_i$ is the volume fraction of the $i$th magnetic component ($\sum_{i} f_i = 1$). The coefficients $2/3$ and $1/3$ account for powder averaging, where $2/3$ of the muon spins precess in internal fields perpendicular (transversal, T) to the field direction, and $1/3$ remain parallel (longitudinal, L) to $B_{{\rm int},i}$.

The nonmagnetic part in ZF experiments is described by the Gaussian Kubo--Toyabe (GKT) function~\cite{kubo1967magnetic}, which can be written as:
\begin{equation}
P_{\rm nm}^{\rm ZF}(t)
  = \frac{2}{3}\left( 1 - \sigma_{\rm GKT}^{2}t^{2}\right)
    e^{\sigma_{\rm GKT}^2 t^2/2}+\frac{1}{3}.
    \label{eq:3}
\end{equation}
Here, $\sigma_{\mathrm{GKT}}$ is the relaxation rate. This function is typically used to describe the nuclear moment contribution in ZF~$\mu$SR experiments.

In WTF experiments, the nonmagnetic part is described by a simple cosine oscillation with exponential relaxation:
\begin{equation}
P_{\rm nm}^{\rm WTF}(t) =  e^{-\lambda\,t}\;\cos \left( \gamma_{\mu} B_{\rm WTF}\; t + \phi \right).
  \label{eq:4}
\end{equation}
Here, $\lambda$ is the exponential relaxation rate, $B_{\rm WTF}$ is the externally applied weak transverse field, and $\phi$ is the initial phase of the muon-spin ensemble.

In the present study, the WTF-$\mu$SR data are analyzed in two modes: the “long” and the “short” time-window modes. In the “long time window” case, the magnetic term in Eq. \eqref{eq:1} was neglected, and the data are analyzed using the nonmagnetic term [Eq. \eqref{eq:4}] only. To avoid the influence of SDW magnetism,
fits started after $\sim 0.3\,\mu\mathrm{s}$ (see solid lines in Fig. \ref{fig:fig3} (\textit{a})). This approach allows tracking the temperature
dependence of the magnetic fraction $f_{\mathrm{m}}$. The short time-window mode considers data up to $1\,\mu\mathrm{s}$ only [Fig.\ref{fig:fig3} (\textit{b})], and the analysis was performed using both the magnetic (m) and nonmagnetic (nm) terms of
Eq. \eqref{eq:1}. This makes it possible to follow the evolution of the strongest component of the internal field ($B_{\rm int,1}$ in our case).

\subsection*{g. The dipole field calculations}

The dipole field calculations were performed assuming the same muon site as reported in \cite{khasanov2024pressure} for both La$_3$Ni$_2$O$_7$ and La$_2$PrNi$_2$O$_7$, and utilized the MUESR code \cite{bonfa2018introduction}.

\section*{V. ACKNOWLEDGEMENTS}

Z.\,G.\ acknowledges support from the Swiss National Science Foundation (SNSF)
through SNSF Starting Grant (No.\ TMSGI2~211750). This work is based on
experiments performed at the Swiss spallation neutron source \textsc{SINQ} and
the Swiss Muon Source \textsc{S$\mu$S}, Paul Scherrer Institute, Villigen,
Switzerland. I.P. acknowledges support from Paul Scherrer Institute research grant No. 2021 01346.

\section*{VI. AUTHOR CONTRIBUTIONS}

IP, RK, and DJG conceived and supervised the project; IP and DJG prepared the samples and performed their X-ray and TG characterization; IP, LK, VP, and SHM performed neutron diffraction experiments and analyzed the data; PFL performed synchrotron diffraction experiments; RK, JJK, HL, ZG, and TJH performed $\mu$SR experiments and analyzed the data; IP, RK, TJH, and DJG wrote the manuscript with contributions from all co-authors. 

\section*{VII. COMPETING FINANCIAL INTERESTS, INCLUSION \& ETHICS}

The authors declare no competing financial interests.
We have read the Journal Portfolio authorship Policy and confirm that this manuscript complies with the policy information about authorship: inclusion \& ethics in global research.

\bibliographystyle{unsrtnat}  
\bibliography{main_arxiv}

\end{multicols}{}

\end{document}